\documentclass[multphys,vecphys]{svmult}
\usepackage{makeidx}      
\usepackage{graphicx}      
\usepackage{multicol}        
\begin{document}

\title*{Scattering Theory of Dynamic Electrical Transport}

\author{M. B\"uttiker\inst{1}\and M. Moskalets\inst{2}}

\institute{D\'epartement de Physique Th\'eorique, Universit\'e de Gen\`eve,
     CH-1211 Gen\`eve 4, Switzerland
\texttt{Markus.Buttiker@physics.unige.ch}
\and Department of Metal and Semiconductor Physics,
     National Technical University "Kharkiv Polytechnic Institute",
     61002 Kharkiv, Ukraine
\texttt{moskalets@kpi.kharkov.ua}}

\maketitle

   \begin{abstract}
We have developed a scattering matrix approach to coherent transport
through an adiabatically driven conductor based on photon-assisted processes.
To describe the energy exchange with the pumping fields we expand 
the Floquet scattering matrix up to linear order in driving
frequency.
\keywords{Emissivity; Instantaneous currents; Internal response; 
Photon--assisted transport; Quantum pump effect; Scattering matrix}
   \end{abstract}

\section{From an Internal Response to a Quantum Pump Effect}
\label{intro}

The possibility to vary several parameters at the same frequency 
but with different phases  \cite{SMCG99} 
of a coherent (mesoscopic) system opens up  
new prospects for the investigation of dynamical quantum transport.
The adiabatic variation of parameters is of particular interest 
since at small frequencies the conductor stays close to an 
equilibrium state: the opening of inelastic conduction channels 
is avoided and quantum mechanical phase coherence is preserved 
to the fullest extend possible.   
The relevant physics has a simple and transparent explanation
within the scattering matrix approach. 

\begin{figure}[t]
\centering
\includegraphics[height=4cm]{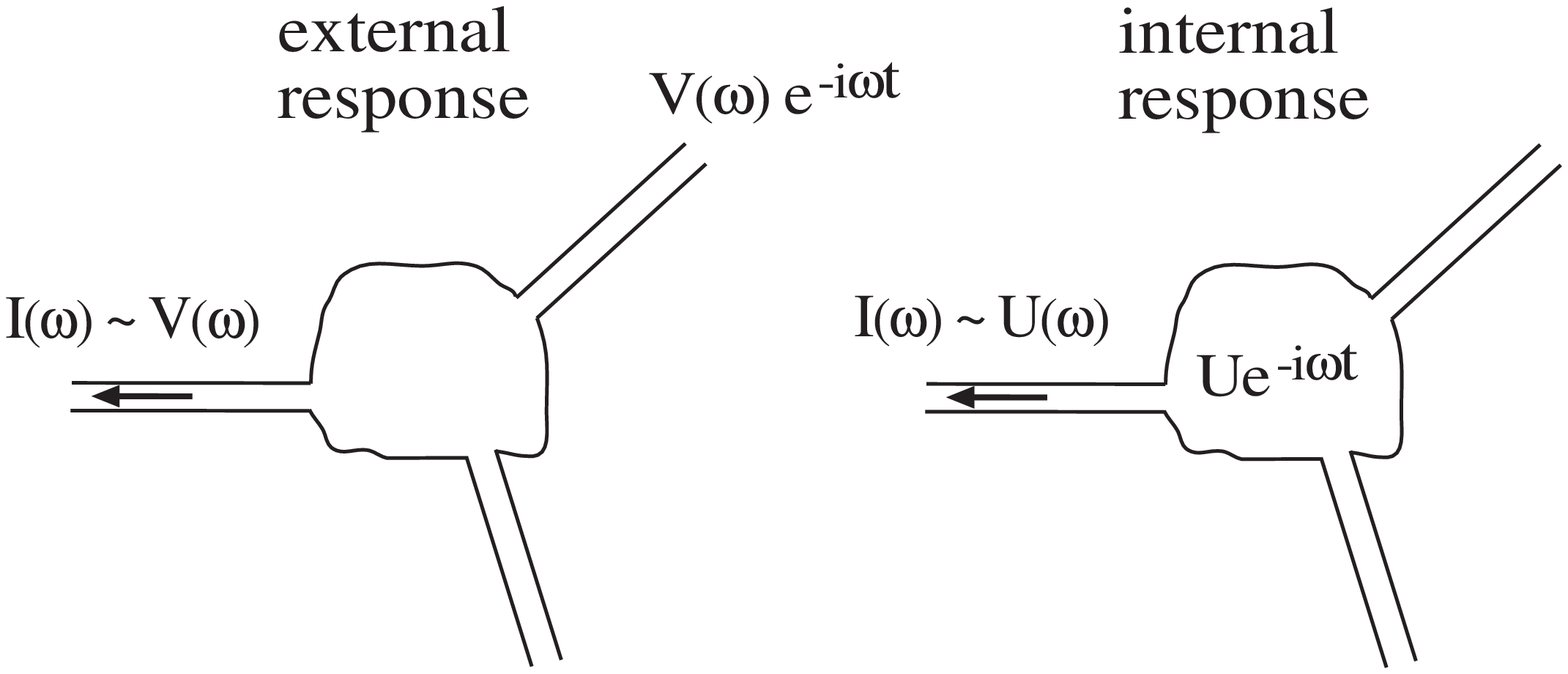}
\caption{External and internal response:
An ac current with amplitude $I(\omega)$, say, in the left lead can
arise as a response either to an oscillating potential 
$V(t)=V(\omega)\E^{-\I\omega t}+V(-\omega)\E^{\I\omega t}$
at one of the external reservoirs, or as a  response to an oscillating potential profile
$\udelta U(\vec{r},t) = \udelta u(\vec{r})
\left( U(\omega)\E^{-\I\omega t}+ U(-\omega)\E^{\I\omega t} \right)$ 
inside the mesoscopic sample.
}
\label{fig1}      
\end{figure}

The variation of parameters leads to a 
dynamic scattering geometry. Quite generally we consider
changes in  the scattering geometry as an {\it internal} response \cite{BTP94}
in contrast to the external response generated by voltages 
applied to the contacts of the conductor, see Fig. \ref{fig1}. 
In general a linear response consists 
both of a response to an external potential oscillations, and 
response to internal potentials. The internal response can be expressed with 
the help of the emissivity $\nu(\alpha,\vec{r})$. 
The emissivity $\nu(\alpha,\vec{r})$ is the portion 
of the density of states at $\vec{r}$ of carriers that will 
exit the conductor through contact $\alpha$. 
The emissivity relates the amplitude $I_{\alpha}(\omega)$ 
of the current in lead $\alpha$
to the amplitude $U(\vec{r}, \omega)$ of a small and slowly oscillating 
internal potential. 
At zero temperature we have \cite{BTP94}
\begin{equation}
\label{Eq1}
I_{\alpha}(\omega) = \I e^2\omega\int \D^3r\nu(\alpha,\vec{r})
U(\vec{r}, \omega).
\end{equation}
Here $e$ is an electron charge and  $\I^{2} = -1$.
The integral in (\ref{Eq1}) runs over the region in which the potential 
deviates from its equilibrium value (typically the volume occupied by the scatterer).

The emissivity is expressed in terms of the scattering matrix $\tens{S}$ 
(of the stationary scatterer)
and its functional derivative with respect to the internal potential variation
$\udelta U(\vec{r},t) = \udelta u(\vec{r})\big(U(\omega)\E^{-\I\omega t} 
+ U(-\omega)\E^{\I\omega t}\big)$:
\begin{equation}
\label{Eq2}
\nu(\alpha,\vec{r}) = -\frac{1}{4\pi\I}\sum_{\beta=1}^\mathrm{N_\mathrm{r}}
\left[ 
S_{\alpha\beta}^{*}\frac{\udelta S_{\alpha\beta}}{\udelta eu(\vec{r})}
- \frac{\udelta S^{*}_{\alpha\beta}}{\udelta eu(\vec{r})}S_{\alpha\beta}
\right].
\end{equation}
The scattering matrix is evaluated at the Fermi energy $E=\mu$.
The summation runs over all the leads (for simplicity here assumed to be single channel) 
connecting the sample to external $N_\mathrm{r} = 1,2,3..$ reservoirs.

Applying the inverse Fourier transformation to (\ref{Eq1}) 
gives the current $I_\alpha(t)$ flowing in response to 
a time-dependent internal potential $U(\vec{r},t)$.
Eq. (1) can easily be generalized to find the response 
to an arbitrary field or parametric variation of the 
scattering geometry \cite{Brouwer98}. 
To arrive at a general expression
we remember that the scattering matrix depends on the internal potential $U$.
That in turn makes $\tens{S}$ time-dependent,
$\tens{S}(t) \equiv \tens{S}[U(t)]$. Thus alternatively we can express 
Eq. (1) in the form 
\begin{equation}
\label{Eq3}
I_{\alpha}(t) = \frac{\I e}{2\pi}\sum_{\beta=1}^\mathrm{N_\mathrm{r}} 
S_{\alpha\beta}^{*}\frac{\D S_{\alpha\beta}}{\D t}.
\end{equation}

Originally in  \cite{BTP94}
the potential $U(\vec{r},t)$ is the self-consistent Coulomb potential.
However the form of (\ref{Eq3}) tells us that 
the current $I_{\alpha}(t)$ can arise in response to a slow variation 
of any quantity (parameter) 
which affects the scattering properties of a mesoscopic sample.
For instance the current generated by a slowly varying 
vector potential \cite{cohen}
permits to derive the Landauer dc-conductance from Eq. (\ref{Eq3}).
But Eq. (\ref{Eq3}) is not limited to the linear response regime.  
The current $I_{\alpha}(t)$ given by Eq. (\ref{Eq3}) 
is a nonlinear functional of the scattering matrix.
Therefore, the mesoscopic system can exhibit an internal rectification effect,
i.e., an oscillating internal potential (or any appropriate oscillating parameter) 
can result in a dc current $I_\mathrm{dc}$. 
Since the elements of the scattering matrix are quantum mechanical amplitudes
the dc-current is the result of a \emph{quantum} rectification process.
This is a quantum pump effect \cite{Brouwer98,cohen,Thouless83,SZBM95,AA98,ZSA99,AEGS00}.
An approach to quantum pumping
based on (1) and (3) was put forth by Brouwer \cite{Brouwer98}. 

Since the time derivative enters (\ref{Eq3}),
quantum rectification will work only 
under special conditions \cite{Brouwer98}.
Let one or several parameters affecting the scattering properties of
a mesoscopic sample change adiabatically and periodically in time.
Then the scattering matrix changes periodically as well. Consider 
the point representing the scattering matrix in the space of all
scattering matrices. 
During the completion of one time period this point 
will move along a closed line ${\cal L}$.
Then the dc current $I_{dc,\alpha}$, 
which is the current $I_{\alpha}(t)$ 
averaged over the time period $T=2\pi/\omega$, 
can be represented as a contour integral in the above mentioned abstract space
\cite{AEGS00}:
\begin{equation}
\label{Eq4}
I_\mathrm{dc,\alpha} = \frac{\I e\omega}{4\pi^2} \oint_{\cal L}
\left( \D \tens{S} \tens{S}^{\dagger}\right)_{\alpha\alpha}.
\end{equation}

The dc current $I_\mathrm{dc,\alpha}$, 
is non-zero if and only if the line ${\cal L}$ encloses a non-vanishing area 
${\cal F}$.
The easy way to see this is to consider a two parameter space
with parameters being $\tens{S}$ and $\tens{S}^\dagger$.
Since $\tens{S}$ and $\tens{S}^{\dagger}$ depend on the same 
set of parameters, 
$\tens{S} = \tens{S}(\{X\mathrm{j}\})$, 
$\tens{S}^{\dagger} = \tens{S}^{\dagger}(\{X\mathrm{j}\}),~
\mathrm{j} = 1,2,\dots, 
N_\mathrm{p}$,
(which includes, for instance, 
the shape of a sample,
the internal potential, 
the magnetic field, 
the temperature, 
the pressure,
the Fermi energy, 
etc.)
then to get the cycle with ${\cal F}\neq 0$
it is necessary to have at least two parameters 
$X_\mathrm{1}(t) = X_\mathrm{1}\cos(\omega t + \varphi_\mathrm{1})$ and 
$X_\mathrm{2}(t) = X_\mathrm{2}\cos(\omega t + \varphi_\mathrm{2} )$
varying with the same frequency $\omega$
and with a phase lag $\varDelta\varphi\equiv\varphi_\mathrm{1}
-\varphi_\mathrm{2}\neq 0$.
In particular, if the oscillating amplitudes are small, 
i.e., if the scattering matrix changes only a little across ${\cal F}$, 
then the pumped current 
is proportional to the square of the cycle area \cite{Brouwer98}:
\begin{equation}
\label{Eq5}
I_\mathrm{dc,\alpha} = \frac{e\omega\sin(\varDelta\varphi)
X_\mathrm{1}X_\mathrm{2}}{2\pi} 
\sum\limits_{\beta=1}^{N_\mathrm{r}}
\Im\left(
\frac{\partial S^{*}_{\alpha\beta}}{\partial X_\mathrm{1}}
\frac{\partial S_{\alpha\beta}}{\partial X_\mathrm{2}}
\right)_\mathrm{X_\mathrm{1}=0,X_\mathrm{2}=0}.
\end{equation}
The very simple and compact expression (4) allows to find the pumped 
current for a wide range of situations. Illustrative examples can be found
in Ref. \cite{AEGS04}.
In the following we will now 
discuss the pumping process from the point of view of 
photon-assisted transport through a mesoscopic system. 
We will show how the interlay between photon--assisted transport and 
quantum mechanical interference results in a quantum pump effect 
\cite{WWG02,MB02}.

\section{Quantum Coherent Pumping: A Simple Picture}
\label{simple}

By nature, the quantum pump effect is a rectification effect.
A single parameter variation only leads to an ac-current. 
In a two parameter variation the modulation of the scatterer
due to the second parameter rectifies the ac-currents generated 
by the first parameter. Rectification is only achieved 
if the driven system can scatter electrons in an asymmetric way. 
Here the asymmetry means that the probability $T_{\alpha\beta}$ 
for an electron to pass through the sample, say, 
from the lead $\beta$ to the lead  $\alpha$
and the probability $T_{\beta\alpha}$ to transit the scatterer in the reverse
direction differ, 
$T_{\alpha\beta} \neq T_{\beta\alpha}$.
Then the flux of electrons entering the scatterer through lead $\alpha$
and the flux of electrons scattered and leaving the system 
through the same lead $\alpha$ differ from each other,
resulting in a net electron flow in lead $\alpha$.
The resulting current can 
be viewed as a result of an asymmetrical redistribution
of incoming flows between the outgoing leads. 

To clarify the physical mechanism which can lead to asymmetric scattering
we now emphasize the essential difference between the driven scatterer and a stationary one. The key difference is the
possibility of photon--assisted transport.
In the stationary case if an electron with energy $E$ enters the phase coherent 
system then it leaves the system with the same energy $E$. 
In contrast, in the driven case while traversing the system
an electron can absorb (or emit) energy quanta $n\hbar\omega$ and
thus it can leave the system with an energy 
$E_\mathrm{n} = E+n\hbar\omega$.

It is important that the electron changes its energy interacting with a system 
which is modulated deterministically. As a consequence the 
inelastic processes is coherent. 
If there are several possibilities for transmission through the system
absorbing or emitting the same energy (say, $n\hbar\omega$)
the corresponding quantum--mechanical amplitudes will interfere.
Such an interference of photon--assisted amplitudes 
can lead to directional asymmetry
of electron propagation through a driven mesoscopic sample.

To illustrate this process we consider a simple but generic example.
It is a system consisting of two regions with oscillating potentials 
$V_\mathrm{1}(t) = 2V\cos(\omega t + \varphi_\mathrm{1})$ and 
$V_\mathrm{2}(t) = 2V\cos(\omega t + \varphi_\mathrm{2})$ 
separated by the distance $L$. 
For the sake of simplicity we assume that both potentials
oscillate with the same small amplitude $2V$.
Consider an electron with energy $E$ incident on the system.  
In leading order in the oscillating amplitudes 
only absorption/emission of a single energy quantum $\hbar\omega$
needs to be taken into account.
So, there are only three scenarios to traverse the system. 
In the first case, an electron does not change its energy, 
the outgoing energy is $E^\mathrm{(out)}=E$.
In the second case, it absorbs one energy quantum,  $E^\mathrm{(out)}=E+\hbar\omega$.
In the third case, it emits an energy $\hbar\omega$, $E^\mathrm{(out)}=E-\hbar\omega$.
Since all these processes correspond to different final states 
(which differ in energy $E^\mathrm{(out)}$ from each other) 
then the full probability $T$
to pass through the system is a sum of three contributions:
\begin{equation}
\label{Eq6}
T = T^\mathrm{(0)}(E;E) + T^\mathrm{(+)}(E+\hbar\omega;E) + 
T^\mathrm{(-)}(E-\hbar\omega;E).
\end{equation}
Here the first argument is an outgoing electron energy
while the second argument is an incoming energy.

The probability $T^\mathrm{(0)}$, like the tranmission probability of stationary scatterer, is insensitive to the propagation direction.
In contrast $T^\mathrm{(+)}$ and $T^\mathrm{(-)}$ are directionally sensitive.
Therefore we concentrate on the last two.

\begin{figure}[t]
\centering
\includegraphics[height=4cm]{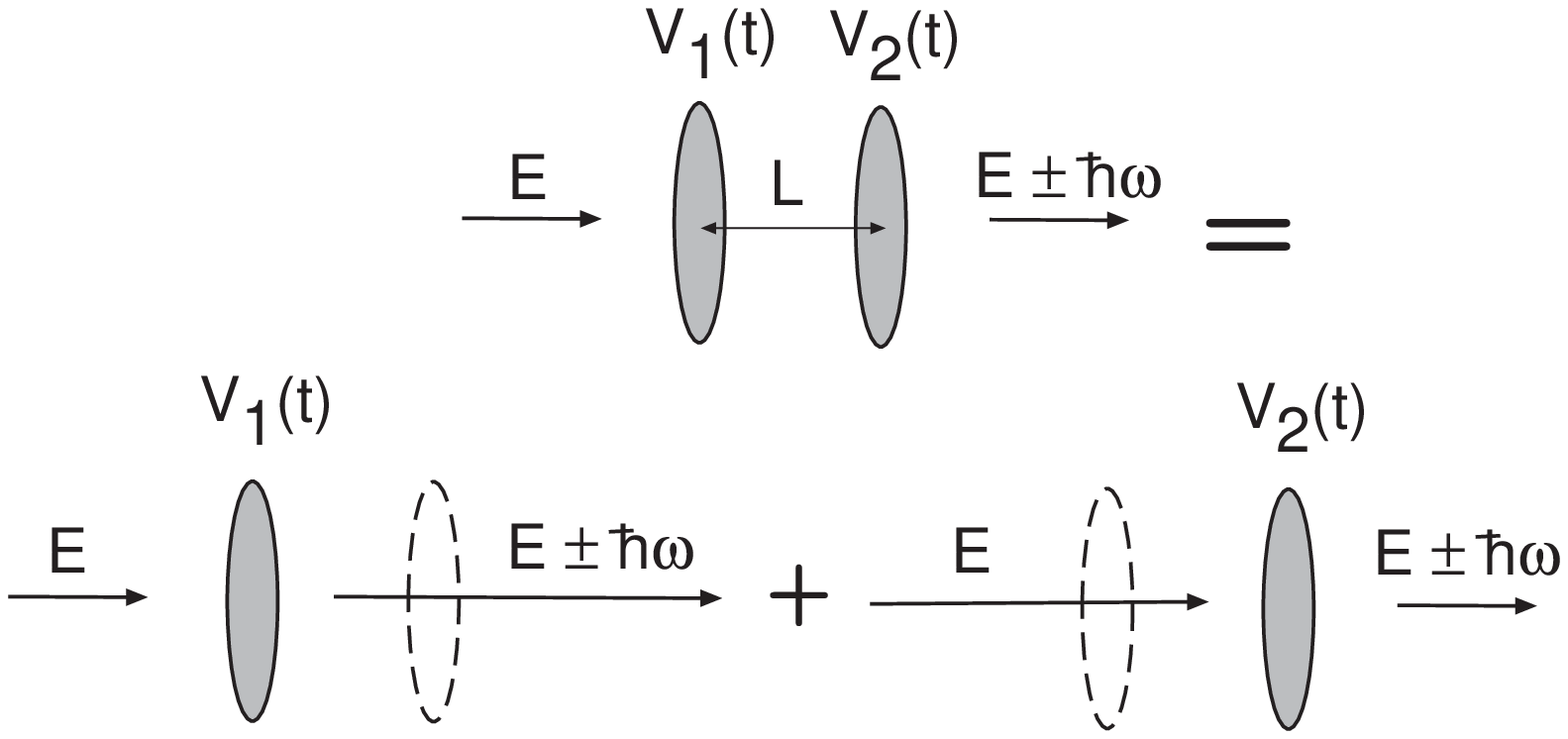}
\caption{For the process in which 
a carrier with energy $E$ traverses two oscillating potentials
$V_\mathrm{1}(t)$ and $V_\mathrm{2}(t)$, a distance $L$ apart,
and 
absorbs/emits an energy quantum $\hbar\omega$
there are
two interfering alternatives. The modulation quantum can be either 
absorbed/emitted at the first barrier or at the second one. }
\label{fig2} 
\end{figure}

Let us consider $T^\mathrm{(+)}$. 
There are two possibilities to pass through the system and
to absorb an energy $\hbar\omega$, see, Fig. \ref{fig2}. 
The first possibility is to absorb an energy due to the oscillation of $V_\mathrm{1}(t)$.
The second possibility is to absorb an energy interacting with 
$V_\mathrm{2}(t)$.
In these two processes an electron 
has the same initial state and the same final state. 
Therefore, we can not distinguish between these two possibilities 
and according to quantum mechanics to calculate the corresponding
probability we first have to add up the corresponding amplitudes
and only then take the square. 
Let us denote the amplitude corresponding to the propagation through the
system with absorbing $\hbar\omega$ at $V_\mathrm{j}$ as 
${\cal A}^\mathrm{(j,+)},~\mathrm{j}=1,2$.
Then the corresponding full probability is:
\begin{equation}
\label{Eq7}
T^\mathrm{(+)} = \left|{\cal A}^\mathrm{(1,+)} 
+ {\cal A}^\mathrm{(2,+)}  \right|^{2}.
\end{equation}
Each amplitude 
(either ${\cal A}^\mathrm{(1,+)}$ or ${\cal A}^\mathrm{(2,+)})$
is a product of two terms, the amplitude 
${\cal A}^\mathrm{(free)}(E)=\E^{\I kL}$ (here $k=\sqrt{2mE}/\hbar$
is an electron wave number)
of free propagation in between the potentials and the amplitude
${\cal A}^\mathrm{(+)}_\mathrm{j}$ 
describing the absorption of an energy quantum $\hbar\omega$
at the potential $V_\mathrm{j}$.
The amplitude ${\cal A}^\mathrm{(+)}_\mathrm{j}$ is proportional to 
the corresponding Fourier coefficient of  $V_\mathrm{j}(t)$. 
The proportionality constant is denoted by $\alpha$.
Therefore we have  ${\cal A}^\mathrm{(+)}_\mathrm{j}=\alpha V_\mathrm{j}
\E^{-\I\varphi_\mathrm{j}}$.

The probability for an electron going from the left to the right
is denoted by $T^\mathrm{(+)}_{\rightarrow}$.
The probability corresponding to the reverse direction -- 
by $T^\mathrm{(+)}_{\leftarrow}$.
Our aim is to show that
\begin{equation}
\label{Eq8}
T^\mathrm{(+)}_{\rightarrow} \neq T^\mathrm{(+)}_{\leftarrow}.
\end{equation}

First we consider $T^\mathrm{(+)}_{\rightarrow}$.
Scattering from the left to the right an electron first meets the potential 
$V_\mathrm{1}(t)$
and only then the potential $V_\mathrm{2}(t)$. 
Therefore if an electron absorbs the energy $\hbar\omega$ at the first potential 
it traverses the remaining part of the system with enhanced energy 
$E_\mathrm{+1} = E + \hbar\omega$. 
The corresponding amplitude is
${\cal A}^\mathrm{(1,+)}_{\rightarrow}=
{\cal A}^\mathrm{(+)}_\mathrm{1} {\cal A}^\mathrm{(free)}(E_\mathrm{+1})$.
While if an electron absorbed $\hbar\omega$ at the second potential
then it goes through the system with the initial energy $E$. 
The quantum mechanical amplitude corresponding to such a process 
reads:
${\cal A}^\mathrm{(2,+)}_{\rightarrow} = 
{\cal A}^\mathrm{(free)}(E){\cal A}^\mathrm{(+)}_\mathrm{2}$.
If the energy quantum is much smaller then the electron energy,
$\hbar\omega \ll E$,
then we can expand the phase factor corresponding to free propagation 
with enhanced energy $E_\mathrm{+1}$
to first order in the driving frequency:
$ k_\mathrm{+1}L \approx \left(k+\frac{\omega}{v}\right)L$,
here $v=\hbar k/m$ is an electron velocity.
Thus we have:
\begin{equation}
\label{Eq9}
\begin{array}{ll}
{\cal A}^\mathrm{(1,+)}_{\rightarrow} \approx 
\alpha V\E^{-\I\varphi_\mathrm{1}}\E^{\I\left(k+\frac{\omega}{v}\right)L}
, \\
\ \\
{\cal A}^\mathrm{(2,+)}_{\rightarrow} =
\E^{\I kL}\alpha V\E^{-\I\varphi_\mathrm{2}}.
\end{array}
\end{equation}
Substituting these amplitudes into (\ref{Eq7}) we obtain the probability
to pass through the system from the left to the right with the absorption of
an energy quantum $\hbar\omega$:
\begin{equation}
\label{Eq10}
T^\mathrm{(+)}_{\rightarrow} = 2\alpha^2V^2\left\{1 + 
\cos\left(\varphi_\mathrm{1} - \varphi_\mathrm{2} 
- \frac{\omega L}{v} \right) \right\}.
\end{equation}
Now we consider the probability $T^\mathrm{(+)}_{\leftarrow}$.
Going from the right to the left an electron first meets the potential 
$V_\mathrm{2}(t)$
and then the potential $V_\mathrm{1}(t)$. 
Therefore, the corresponding amplitudes are:
\begin{equation}
\label{Eq11}
\begin{array}{ll}
{\cal A}^\mathrm{(1,+)}_{\leftarrow} = 
\E^{\I kL}\alpha V\E^{-\I\varphi_\mathrm{1}}.
, \\
\ \\
{\cal A}^\mathrm{(2,+)}_{\leftarrow} \approx 
\alpha V\E^{-\I\varphi_\mathrm{2}}\E^{\I\left(k+\frac{\omega}{v}\right)L}
\end{array}
\end{equation}
Using  (\ref{Eq7}) and  (\ref{Eq11}) we find:
\begin{equation}
\label{Eq12}T^\mathrm{(+)}_{\leftarrow} = 2\alpha^2V^2\left\{1 + 
\cos\left(\varphi_\mathrm{1} - \varphi_\mathrm{2} 
+ \frac{\omega L}{v} \right) \right\}.
\end{equation}
Comparing (\ref{Eq10}) and (\ref{Eq12}) we see
that the transmission probability depends on the direction of
electron propagation 
as we announced in (\ref{Eq8}).

Let us characterize the asymmetry in transmission probability by the difference
$\Delta T^\mathrm{(+)} = T^\mathrm{(+)}_{\rightarrow} 
- T^\mathrm{(+)}_{\leftarrow}$:
\begin{equation}
\label{Eq13}
\Delta T^\mathrm{(+)} = 4\alpha^2V^2\sin(\varDelta\varphi)
\sin\left( \frac{\omega L}{v} \right).
\end{equation}
In our simple case the emission leads to the same asymmetry in 
the photon--assisted transmission probability: 
$\varDelta T^\mathrm{(-)} = \varDelta T^\mathrm{(+)}$.
Therefore if there are the same electron flows $I_\mathrm{0}$ 
coming from the left and 
from the right, then a net current 
$I = I_\mathrm{0}\left(\varDelta T^\mathrm{(-)}
+ \varDelta T^\mathrm{(+)}\right)$ 
is generated.
We assume a positive current to be directed to the right.

We see that the induced current $I$ depends separately on 
$\varDelta\varphi = \varphi_\mathrm{1} - \varphi_\mathrm{2}$
and on the factor $\omega L/v$. This is an additional dynamical phase 
due to absorption of an energy quantum $\hbar\omega$.
Such a separation of phase factors 
can be interpreted in the following way.
The presence of the phase lag $\varDelta\varphi$ (by modulo $2\pi$)
between the oscillating potentials
$V_\mathrm{1}(t)$ and $V_\mathrm{2}(t)$ 
breaks the time reversal invariance of a problem
and hence potentially permits the existence of a steady particle flow.
The second term emphasizes a spatial asymmetry of a model
consisting of two inequivalent oscillating regions separated by a distance $L$,
and tells us that the interference of photon--assisted amplitudes
is the mechanism inducing electron flow. Neither a phase lag nor a photon-assisted
process can separately lead to a dc current.

The simple model presented here reproduces several generic properties 
of a periodically driven quantum system. 
First, a driven spatially asymmetric system 
can pump a current 
between external electron reservoirs to which this system is coupled.
Second,  the pumped current is periodic in the phase lag
between the driving parameters.
Third, at small driving frequency, $\omega\to 0$,
the current generated is linear in $\omega$. In contrast
the oscillatory  dependence on $\omega$ of (\ref{Eq13})
is a special property of the simple resonant tunneling structure \cite{MBstrong02}.

\section{Beyond the Frozen Scatterer Approximation:\\ Instantaneous currents}
\label{ic}

In experiments 
the external electrical circuit
to which the pump is connected is 
important \cite{polian,vavilov}. Consequently, 
it is of interest to understand the workings of a quantum pump connected 
to contacts which are not at equilibrium but support dc-voltages and 
ac-potentials. We now present a number of results of a rigorous calculation of 
dynamically generated currents within the scattering matrix approach
for spinless non-interacting particles \cite{MBac03}. These results permit
the investigation of pumping also in experimentally more realistic 
non-ideal situations. 

We generalize the approach of  \cite{BTP94} to the case of 
strong periodic driving when many photon processes are of importance.
To take them into account we use the Floquet scattering matrix 
whose elements $S_\mathrm{F,\alpha\beta}(E_\mathrm{n},E)$ 
are the quantum mechanical amplitudes 
(times $\sqrt{k_\mathrm{n}/k}$, where 
$k_\mathrm{n}=\sqrt{2mE_\mathrm{n}}/\hbar$,
$E_\mathrm{n} = E+n\hbar\omega$)
for scattering of an electron with energy $E$
from lead $\beta$ to lead $\alpha$ with absorption ($n>0$) or emission 
($n<0$) of $|n|$ energy quanta $\hbar\omega$.
Like in  \cite{BTP94}
we deal with low frequencies (adiabatic driving)
and calculate the current linear in $\omega$.
In this limit we can expand the Floquet scattering matrix 
in powers of $\omega$. 

To zero-th order the Floquet sub-matrices 
$\tens{S}_\mathrm{F}(E_\mathrm{n},E)$ are merely
the matrices of Fourier coefficients $\tens{S}_\mathrm{n}$ 
of the stationary scattering matrix 
with time-dependent (pump)-parameters,
$\tens{S}(t) \equiv \tens{S}(\{X_\mathrm{j}(t)\})$. 
The matrix $\tens{S}(t)$ is the 
{\it frozen} scattering matrix, in the sense that it describes
the time moment $t$ and hence stationary scattering.
If all the reservoirs are kept at the same conditions (potential, temperature..) then the knowledge
of only the frozen scattering matrix is sufficient to calculate the current 
flowing through the scatterer, see (\ref{Eq3}).
Under more general conditions knowledge of the frozen scattering matrix 
is not sufficient.
We stress that the frozen scattering matrix does not describe
the scattering of electrons by a time-dependent scatterer: only the Floquet
scattering matrix does.
Since we are interesting in a current linear in $\omega$ then, in general, 
it is necessary to know the Floquet scattering matrix with the same accuracy.
Thus we have to go beyond the frozen scattering matrix approximation and
to take into account the corrections of order $\omega$.
As we illustrated in Sec.\ref{simple} such corrections are due to
interference between photon--assisted amplitudes.

We use the following ansatz:
\begin{equation}
\label{Eq14}
\tens{S}_\mathrm{F}(E_\mathrm{n},E) = \tens{S}_\mathrm{n}(E)  
+ \frac{n\hbar\omega}{2}\frac{\partial\tens{S}_\mathrm{n}(E)}{\partial E}  
+ \hbar\omega\tens{A}_\mathrm{n}(E) + O(\omega^2).  
\end{equation}
The matrix $\tens{A}(E,t)$ 
[with Fourier coefficients $\tens{A}_\mathrm{n}(E)$] introduced here
is a key ingredient.  
As we will see this matrix reflects the asymmetry in scattering
from one lead to the other and back. 
The unitarity condition for the Floquet scattering matrix leads to the 
following equation for the matrix $\tens{A}(E,t)$: 
\begin{equation}
\label{Eq15}
\hbar\omega\left(\tens{S}^{\dagger}(E,t)\tens{A}(E,t) 
+ \tens{A}^{\dagger}(E,t)\tens{S}(E,t)\right)
= \frac{1}{2}{\cal P}\{\tens{S}^{\dagger};\tens{S} \},
\end{equation}
where ${\cal P}$ is the Poisson bracket with respect to energy and time 
$$
{\cal P}\{\tens{S}^{\dagger};\tens{S} \} =
\I\hbar \left( \frac{\partial \tens{S}^{\dagger}}{\partial t}
\frac{\partial \tens{S}}{\partial E} -
\frac{\partial \tens{S}^{\dagger}}{\partial E}
\frac{\partial \tens{S}}{\partial t}
\right).
$$
The matrix $\tens{A}$ can not be expressed in terms of 
the frozen scattering matrix $\tens{S}(t)$ 
and it has to be calculated (like $\tens{S}$ itself) in each particular case.
Nevertheless there are several advantages in using (\ref{Eq14}).

First, the matrix $\tens{A}$ has a much smaller number of  elements 
than the Floquet scattering matrix.
The matrix $\tens{A}$ depends on only one energy, $E$, and, therefore,
it has $N_\mathrm{r}\times N_\mathrm{r}$ 
elements like the stationary scattering matrix $\tens{S}$.
In contrast, the Floquet scattering matrix $\tens{S}_\mathrm{F}$ 
depends on two energies, $E$ and $E_\mathrm{n}$, and, therefore,
it has $(2n_\mathrm{max}+1)\times 
N_\mathrm{r}\times N_\mathrm{r}$ relevant elements. 
Here $n_\mathrm{max}$ 
is the maximum number of energy quanta $\hbar\omega$
absorbed/emitted by an electron interacting with the scatterer 
which we have to take into account to correctly describe the scattering process.
For small amplitude driving we have $n_\mathrm{max}\approx 1$, 
whereas if the parameters vary with a large amplitude then 
$n_\mathrm{max}\gg 1$.

Second, the Floquet scattering matrix has no definite symmetry with respect
to a magnetic field $H$ reversal. In contrast both the frozen scattering
matrix $\tens{S}$ and  $\tens{A}$ have. 
The analysis of the micro-reversibility of the equations of motion gives 
the following symmetry:
\begin{equation}
\label{Eq16}
\begin{array}{l}
\tens{S}(-H) = \tens{S}^\mathrm{T}(H), \\
\ \\
\tens{A}(-H) = - \tens{A}^\mathrm{T}(H),
\end{array}
\end{equation}
where the upper index "$\mathrm{T}$" denotes the transposition.
In the absence of a magnetic field, $H=0$, 
the matrix $\tens{A}$ is antisymmetric in lead indices,
$A_{\alpha\beta} = - A_{\beta\alpha}$. 

Next, using the adiabatic expansion, (\ref{Eq14}),
we calculate the full time-dependent current $I_{\alpha}(t)$ 
flowing in lead $\alpha$ as follows \cite{MBac03}:
\begin{equation}
\label{Eq17}
\begin{array}{c}
I_{\alpha}(t) =    
\int\limits_{0}^{\infty} \D E  
\sum\limits_{\beta}  
\bigg\{   
\frac{e}{h} 
\left[ f_\mathrm{0,\beta}- f_\mathrm{0,\alpha}\right]   
\big|S_{\alpha\beta}(E,t)\big|^2  \\
\ \\
- e \frac{\partial}{\partial t} 
 \left[ f_\mathrm{0,\beta}\frac{\D N_{\alpha\beta}(E,t)}{\D E}\right] 
+ f_\mathrm{0,\beta} \frac{\D I_{\alpha\beta}(E,t)}{\D E}
\bigg\}. 
\end{array}
\end{equation}
Here $f_\mathrm{0,\alpha}$ is the Fermi distribution function for electrons
in reservoir $\alpha$.  
We assume a current in a lead to be positive if it is directed
from the scatterer to the corresponding reservoir.
Equation (\ref{Eq17}) generalizes (\ref{Eq3}) 
to the case with external reservoirs being nonidentical
(e.g., having different chemical potentials, temperatures, etc.).
The three parts in the curly brackets on the RHS of (\ref{Eq17}) 
can be interpreted as follows: 
The first part defines the currents injected from the external reservoirs. 
It depends on the time-dependent conductance 
$G_{\alpha\beta}(t) = \frac{e^2}{h}|S_{\alpha\beta}(t)|^2$ 
of the frozen scatterer and hence it describes a classical rectification
contribution to the dc current $I_\mathrm{dc,\alpha}$.
The second part defines the current generated
by the oscillating charge $Q(t)$ of the scatterer:
\begin{equation}
\label{Eq18}
Q(t) = e \sum\limits_{\alpha} \sum\limits_{\beta}  
\int\limits_{0}^{\infty} \D E f_\mathrm{0,\beta} 
\frac{\D N_{\alpha\beta}(E,t)}{\D E}. 
\end{equation}
Here $\D N_{\alpha\beta}/\D E$ is the global partial density of states 
for a frozen scatterer:
\begin{equation}
\label{Eq19}
\frac{\D N_{\alpha\beta}}{\D E} = \frac{\I}{4\pi} 
\left(  
\frac{\partial S^{*}_{\alpha\beta}}{\partial E} S_{\alpha\beta} 
- S^{*}_{\alpha\beta}\frac{\partial S_{\alpha\beta}}{\partial E}  
\right). 
\end{equation}
Apparently this part gives no contribution to the dc current.
The third part describes the currents generated by the oscillating scatterer.
The ability to generate these ac currents differentiates a non-stationary 
dynamical scatterer from a merely frozen scatterer. 

The \emph{instantaneous spectral currents} $\D I_{\alpha\beta}/\D E$
pushed by the oscillating scatterer from lead $\beta$ to lead $\alpha$
read:
\begin{equation}
\label{Eq20}
\frac{\D I_{\alpha\beta}}{\D E} = \frac{e}{h}\Big(
2\hbar\omega Re[S^{*}_{\alpha\beta}A_{\alpha\beta}]
+\frac{1}{2}{\cal P}\{S_{\alpha\beta}; S^{*}_{\alpha\beta}\} \Big). 
\end{equation} 
The two terms in this equation have different symmetry properties
with respect to the interchange of lead indices. 
That is most evident in the absence of a magnetic field, $H=0$.
In this case the first term on the RHS of (\ref{Eq20})
is antisymmetric in lead indices while the second term 
has the symmetry of the stationary scattering matrix $\tens{S}$ and is 
thus symmetric.
Therefore, the matrix $\tens{A}$ is responsible for the 
directional asymmetry of dynamically generated currents:
\begin{equation}
\label{Eq21}
\frac{\D I_{\alpha\beta}}{\D E} \neq \frac{\D I_{\beta\alpha}}{\D E}.
\end{equation}

We remark, that if one calculates the dc current generated
in the particular case when all the reservoirs are at same potentials
and temperatures ($f_\mathrm{0,\alpha} = f_\mathrm{0,\beta},
~\forall~\alpha,\beta$)
then (\ref{Eq17}) (after averaging over a pump period) 
generates Brouwer's result, (\ref{Eq4}).
In this case the matrix $\tens{A}$ plays no role. 
Different contributions of this matrix can be combined 
as in (\ref{Eq15}). In contrast, the matrix $\tens{A}$ is important
in less symmetrical situations, when the electron flows arriving at the
scatterer from different leads are different. 

\begin{figure}[t]
\centering
\includegraphics[height=4cm]{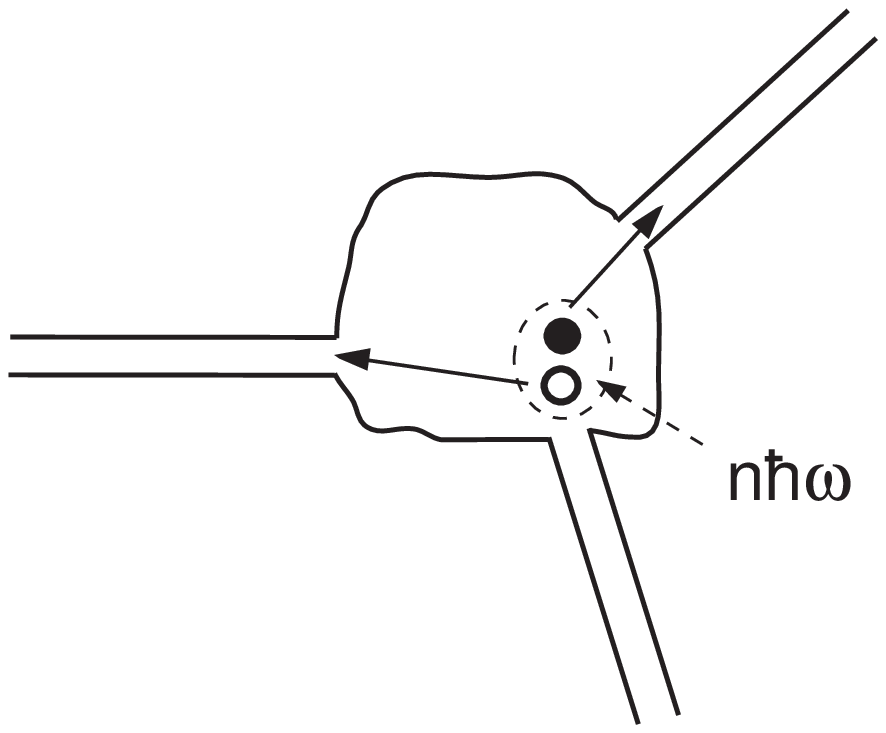}
\caption{
Interacting with an oscillating potential
the electron system gains modulation quanta of energy.
Absorption of an energy $n\hbar\omega$ leads to the creation 
of a non-equilibrium quasi--electron--hole pair.
The quasi--electron (black circle) and hole (open circle)
can leave the scattering region through different leads.
This leads to current pulses of different signs in 
the corresponding leads.
}
\label{fig3} 
\end{figure}

The spectral currents $\D I_{\alpha\beta}/\D E$ are subject to the following 
conservation law: 
\begin{equation}
\label{Eq22}
\sum\limits_{\alpha=1}^{N_\mathrm{r}} 
\frac{\D I_{\alpha\beta}(E,t)}{\D E} = 0.
\end{equation}

Such a property supports the point of view that these currents arise 
inside the scatterer.
They are generated by the non-stationary scatterer  
without any external current source.  
The appearance of currents 
subject to the conservation law (\ref{Eq22})
can be easily illustrated within the quasi--particle picture \cite{MB02}.
Let all the reservoirs have the same chemical potential 
$\mu_\alpha = \mu,~\alpha=1, \dots, N_\mathrm{r}$. 
We introduce quasi--particles: 
the quasi--electrons corresponding to filled states with energy $E>\mu$ and 
holes corresponding to empty states with energy $E<\mu$. 
Then at zero temperature there are no incoming quasi--particles.
In other words, from each lead the vacuum of quasi--particles 
is falling upon the scatterer.
Interacting with the oscillating scatterer, the system of (real) electrons can gain, 
say, $n$ energy quanta $\hbar\omega$. 
In the quasi--particle picture this process corresponds to the creation of 
a quasi--electron--hole pair with energy $n\hbar\omega$.
The pair dissolves and the quasi--particles are scattered separately
to the same or different leads, see, Fig. \ref{fig3}. 
If the scattering matrix depends on energy then the quasi--electron and hole
are scattered, on average, into different leads since they 
have different energies.
Suppose the electron leaves the scattering region through lead $\alpha$
and the hole leaves through lead $\beta$.
Since electrons and holes have opposite charge the current pulses
created in the leads $\alpha$ and $\beta$ have different sign.
As a result a current pulse arises between the $\alpha$ and $\beta$ reservoirs.
In this picture it is 
evident that there is no incoming current and the sum of outgoing 
currents does satisfy the conservation law (\ref{Eq22}).

Note, from (\ref{Eq15}) and (\ref{Eq20}) it is obvious that 
a conductor with strictly
energy independent scattering matrix does not produce current
and thus it does not show a quantum pump effect.

The current $I_{\alpha}(t)$, calculated with (\ref{Eq17}), satisfies
the continuity  equation:
\begin{equation} 
\label{Eq23} 
\sum\limits_{\alpha}I_{\alpha}(t) +  \frac{\partial Q(t)}{\partial t} = 0, 
\end{equation} 
and thus conserves charge.
To demonstrate this we use the unitarity of the frozen scattering matrix,
$\sum_{\alpha} |S_{\alpha\beta}|^2 = \sum_{\beta}  |S_{\alpha\beta}|^2 =1$,
and the definition of the charge $Q(t)$ of the scatterer, see  (\ref{Eq18}).

It follows from (\ref{Eq22}) that
the dynamically generated currents $\D I_{\alpha\beta}/\D E$
do not contribute to (\ref{Eq23}).
Therefore they have nothing to do with charging/discharging of a scatterer.
The existence of these currents is an intrinsic property of dynamical scatterer. 

In conclusion, 
we clarified the role played by photon--assisted processes in 
adiabatic electron transport through a periodically driven mesoscopic system. 
The interference between the corresponding photon--assisted amplitudes makes 
the transmission probability dependent on the electron transmission direction 
in striking contrast with stationary scattering.
To consider properly this effect we 
introduced an adiabatic (in powers of a driving frequency) expansion of
the Floquet scattering matrix and demonstrated that already
linear in $\omega$ terms exhibit the required asymmetry.

The ability to generate currents is only one interesting aspect of 
quantum pumps. Recent works point to the possibility of dynamical 
controlled generation of entangled electron-hole states 
\cite{samuelsson,beenakker}.  This brings into focus the dynamic 
quantum state of pumps. This and other properties will 
likely assure a continuing lively interest in quantum pumping.


\end{document}